\documentclass[aps,prc,twocolumn,superscriptaddress]{revtex4-2}
\usepackage{CJKutf8}
\usepackage{amssymb}
\usepackage{amsmath}
\usepackage{diagbox}
\usepackage{makecell}
\usepackage{tikz}
\usepackage{color}
\usepackage{graphicx}
\usepackage{multirow} 
\usepackage{makecell}
\usepackage{ulem}
\usepackage[hidelinks]{hyperref}
\usepackage{url}
\usepackage{rotating} 
\usepackage{threeparttable} % 添加这个包用于更好的表格脚注支持
\usepackage{booktabs} % 添加这个包用于更好的表格线控制
\usepackage{longtable}  
\usepackage{array} 
\usepackage{bm}
\usepackage{supertabular}
\usepackage{array} % 用于精确控制列宽
\usepackage{ragged2e}
\begin{document}

%\title{ Research Progress on $^{229}$Th from the Perspective of Nuclear Physics }
\title{Flat ``Island of Stability'' of Superheavy Nuclei}

\author{Yi-Ming Jiang}
\affiliation{Institute of Theoretical Physics, Chinese Academy of Sciences, Beijing 100190, China}
\affiliation{School of Physical Sciences, University of Chinese Academy of Sciences, Beijing 100049, China}

\author{Xiao Lu }
\affiliation{Institute of Theoretical Physics, Chinese Academy of Sciences, Beijing 100190, China}

\author{T. M. Shneidman}
\affiliation{Joint Institute for Nuclear Research, 141980 Dubna, Moscow Region, Russia}

\author{N. V. Antonenko}
\affiliation{Joint Institute for Nuclear Research, 141980 Dubna, Moscow Region, Russia}

\author{Shan-Gui Zhou }
\email[]{sgzhou@itp.ac.cn}
\affiliation{Institute of Theoretical Physics, Chinese Academy of Sciences, Beijing 100190, China}
\affiliation{School of Physical Sciences, University of Chinese Academy of Sciences, Beijing 100049, China}
\affiliation{School of Nuclear Science and Technology, University of Chinese Academy of Sciences, Beijing 100049, China}

%\thanks{} 
\date{\today}

\begin{abstract}

We present the first systematic microscopic calculations of fission barrier heights for 292 even-even superheavy nuclei using the multidimensionally constrained relativistic mean field model with triaxial and octupole deformations treated simultaneously. The calculated barrier heights remain around 5 MeV  in the vicinity of the predicted doubly magic nucleus ${}^{304}120$. This flat landscape contrasts with the conventional picture of mutually enhanced magicity and indicates that the proton shell closure at $Z=120$ and the neutron shell closure at $N=184$ do not reinforce but suppress each other. We introduce an $\alpha$-linked energy difference $\Delta_{\rm pn}$ to identify the mutually suppressed magicity caused by distinct proton and neutron density distributions. These results challenge the expected shell landscape of superheavy nuclei and may have a crucial impact on estimating survival probabilities in the synthesis of superheavy nuclei.

\end{abstract}

%\keywords{}

\maketitle
The synthesis of new elements has long attracted great interest in modern science \cite{Sanders2021}. Over the past two decades, experimental efforts using fusion-evaporation reactions have extended the periodic table to $Z = 118$, completing the seventh period \cite{Oganessian2015_RPP78_036301,Hofmann2015_JPG42_114001,Smits2023_NP6_86,Oganessian2024_EPJA60_227,Jiang2025_PHYSICS54-599}. The synthesis of elements 119 and 120 is currently under way at several heavy-ion accelerator facilities worldwide \cite{Hofmann2016_EPJA52_180,GIULIANI_RMP91_011001,Zhang2025_NST36_204}. However, the fusion-evaporation cross sections are extremely small, typically yielding only a few detected events after months of beam time. As a result, experimental information for superheavy nuclei is largely limited to $\alpha$-decay properties, while quantities such as fission barrier heights are difficult to extract from the available experimental data. Fission barrier heights are of particular importance, because they characterize the stability of superheavy nuclei against fission and serve as key inputs for estimating the survival probabilities of compound nuclei in fusion-evaporation reactions \cite{Adamian2000_PRC62_064303,Xia2011_SCPMA}. This situation thus calls for reliable systematic theoretical calculations of fission barriers.

% \sout{This situation thus calls for reliable systematic theoretical calculations of fission barriers. }\red{(后面自然会引出理论缺口，这句过渡有点空，建议删掉)}    

The distribution of fission barrier heights across the nuclear chart of superheavy nuclei reflects the landscape of the ``island of stability'' and probes the underlying shell effects. In known nuclear regions, shell effects are mutually reinforced near doubly magic nuclei, a phenomenon referred to as mutually enhanced magicity (MEM) \cite{SCHMIDT1979_NPA318_253,ZELDES1983_NPA399_11,Lunney2003_RMP75_1021,Vladimir2023_EPJA59_22}. For superheavy nuclei, enhanced stability against fission has long been expected around the next doubly magic nucleus \cite{SOBICZEWSKI1966_PL_22_500,Oganessian2015_RPP78_036301,Hofmann2015_JPG42_114001,Hofmann2016_EPJA52_180,GIULIANI_RMP91_011001,Smits2023_NP6_86,Oganessian2024_EPJA60_227,Zhang2025_NST36_204,Jiang2025_PHYSICS54-599}. If MEM persists in this region, the fission barrier landscape should exhibit a pronounced peak near this nucleus. 

% \sout{However, the nuclear structure in this region is extremely complex. Therefore, systematic investigation of fission barriers and the resulting landscape is essential for understanding shell evolution in superheavy nuclei and for guiding the synthesis of new elements.} \red{(这两句话比较泛，删掉如何？)}

Fission barriers of superheavy nuclei have been studied with the extended Thomas-Fermi plus Strutinsky integral (ETFSI) method \cite{MAMDOUH2001_NPA679_337}, macroscopic-microscopic (MM) models \cite{MOLLER2015_PRC91_024310,Jachimowicz_PRC95_014303,JACHIMOWICZ2021_ADNDT138_101393,Pomorski2018,Pomorski2022,Wang2024_CPC48_094103}, and microscopic models primarily based on nuclear density functional theory (DFT) \cite{Egido2000_PRL85_1198,Zhang2003_CPL20_1694,Burvenich2004_PRC69_014307,Samyn2005_PRC72_044316,Pei2009_PRL102_192501,Warda2012_PRC86_014322,Staszczak2013_PRC87_024320,Abusara2012_PRC85_024314,Agbemava2017_PRC_95_054324,Shi2019_PRC_99_064316,Taninah2020_PRC102_054330,MARDYBAN2026_PRC113_054308}. Systematic MM calculations have provided barrier inputs for reaction models and generally predict a peaked island landscape \cite{MOLLER2015_PRC91_024310,Jachimowicz_PRC95_014303,JACHIMOWICZ2021_ADNDT138_101393,Malov2021_PRC104_L011304,Wang2024_CPC48_094103}. Microscopic calculations offer a more self-consistent description of nuclear structure and fission properties, but relevant studies have usually focused on selected nuclei or imposed axial or reflection symmetry in global surveys. Since triaxial and octupole deformations have been shown to be important for the structure and potential energy surfaces of superheavy nuclei \cite{CWIOK2005_Nature433_705,Abusara2012_PRC85_024314,Meng2019_SCPMA63_212011,Agbemava2016_PRC93_044304,Jachimowicz2017_PRC95_034329,Egido2020_PRL125_192504,Egido2021_PRL126_192501,Xu2024_PRL133_022501,MARDYBAN2026_PRC113_054308}, a large-scale microscopic study of fission barriers including
triaxial and octupole deformations is warranted.

The multidimensionally constrained relativistic mean field (MDC-RMF) model has been developed to describe nuclear potential energy surfaces with multiple shape degrees of freedom \cite{Lu2012_PRC85_011301,Lu2014_89_014323,Zhou_2016,Zhao2017_PRC95_014320}. Its applications to actinide fission barriers provide a basis for extending the same framework to superheavy nuclei \cite{Lu2012_PRC85_011301,Lu2014_89_014323,Goriely2025RIPL}.

% The multi-dimensionally constrained relativistic mean field (MDC-RMF) model \cite{Lu2012_PRC85_011301,Lu2014_89_014323,Zhou_2016,Zhao2017_PRC95_014320} has been developed to study various shapes and their manifestations in nuclei. As MDC-RMF has been successfully applied to the fission barriers of actinides \cite{Lu2012_PRC85_011301,Lu2014_89_014323}, it is natural to extend this approach to study superheavy nuclei.

In this Letter, we present the first systematic microscopic calculation of fission barrier heights for 292 even-even superheavy nuclei within the MDC-RMF model with triaxial and octupole deformation treated simultaneously. The results show that the fission barrier heights remain around 5 MeV near the doubly magic nucleus ${}^{304}120$, indicating a flat landscape in the ``island of stability''. We point out that this anomaly arises from mutually suppressed magicity (MSM) in superheavy nuclei, contrasting with the conventional MEM in known nuclei \cite{Lunney2003_RMP75_1021}, and reveal the microscopic origin of this effect.

The MDC-RMF model starts from the Lagrangian of covariant density functional theory \cite{Serot1984_ANP16_1,Reinhard1989_RPP52_439,RING1996_PPNP36_193,VRETENAR2005_PR409_101,MENG2006470,Paar_2007,NIKSIC2011519,Zhou_2016}, and within the mean field approximation, the Dirac equation for nucleons reads 
\begin{equation}
	\{\boldsymbol{\alpha}\cdot\boldsymbol{p}+\beta[M+S(\boldsymbol{r})]+V(\boldsymbol{r})\}\psi_k(\boldsymbol{r})=\varepsilon_k\psi_k(\boldsymbol{r}),
\end{equation}
where the vector potential $V(\boldsymbol{r})$ and scalar potential $S(\boldsymbol{r})$ are determined self-consistently from the densities. To deal with pairing correlations, the BCS method is used with a separable pairing force of finite range \cite{Tian2006_CPL23_3226,Tian2009_PRC79_064301,TIAN2009_PLB676_44,Tian2009_PRC80_024313}: 
\begin{equation}
V_{\rm pair}
=-G\delta(\boldsymbol{R}-\boldsymbol{R}')
P(\boldsymbol{r})P(\boldsymbol{r}')\frac{1-P^\sigma}{2},
\end{equation}
where $\boldsymbol{R}=(\boldsymbol{r}_1+\boldsymbol{r}_2)/2$ and $\boldsymbol{r}=\boldsymbol{r}_1-\boldsymbol{r}_2$ are the center-of-mass and relative coordinates, respectively. The factor $P(\boldsymbol{r})$ is a Gaussian function, $P(\boldsymbol{r})=(4\pi a^2)^{-3/2}\exp(-r^2/4a^2)$, and $P^\sigma$ is the spin-exchange operator. The separable pairing force is of finite range and translational invariance, which avoids artificial cutoffs, and reproduces with high accuracy the nuclear-matter pairing gap obtained with the Gogny forces at a reduced computational cost.
% \sout{where $\boldsymbol{R}=(\boldsymbol{r}_1+\boldsymbol{r}_2)/2$ and $\mathbf{r=r_1-r_2}$ are center-of-mass and relative coordinates, respectively. The factor $P(r)$ is a Gaussian function $P(r)=(4\pi a^2)^{-3/2}\exp{\left(-r^2/4a^2\right)}$. }  

To calculate the potential
energy surfaces, we use the modified linear constraint method for the deformation parameters,
\begin{equation}
E' = E_{\rm RMF}
+ \sum_{\lambda\mu} \frac{1}{2} C_{\lambda\mu}\beta_{\lambda\mu},
\end{equation}
where $C_{\lambda\mu}$ are constraint coefficients updated during the iteration, and $\beta_{\lambda\mu}$ are deformation parameters defined from the corresponding multipole moments.
% %defined as
% \begin{equation}
% 	\beta_{\lambda\mu}=\frac{4\pi}{3AR^\lambda}\langle \hat{Q}_{\lambda\mu} \rangle = \frac{4\pi}{3AR^\lambda}\int\rm{d}^3\bold{r}\rho_Vr^\lambda Y_{\lambda\mu}(\theta,\phi).
% \end{equation}
% where $Y_{\lambda\mu}(\theta,\phi)$ are spherical harmonics and $\rho_V$ is vector density. 
The Dirac equation is solved by expanding spinors in the  axially deformed harmonic oscillator basis \cite{GAMBHIR1990_AoP198_132,RING1997_CPC105_77}, with deformation parameters compatible with $V_4$ symmetry, such as $\beta_{20}$, $\beta_{22}$, and $\beta_{30}$, included self-consistently. More details of the MDC-RMF model can be found in Refs.~\cite{Lu2012_PRC85_011301,Lu2014_89_014323,Zhou_2016}.

Our calculations of fission barrier heights are carried out with the covariant density functional PC-PK1 \cite{ZHAO2010_PRC82_054319,ZHAO2012_PRC86_064324}. The pairing parameters  are $G_0 = 728$ MeV fm${}^{3}$ and $a=0.644$ fm \cite{TIAN2009_PLB676_44}. The scaling factor $G/G_0=1.08$ is adjusted to odd-even mass staggering near $^{208}$Pb and $^{270}$Hs from experimental and estimated masses \cite{Huang2021_CPC45_030002,Wang2021_CPC45_030003}. The basis is truncated at $N_{\rm F}=20$ major shells for the Dirac spinors. The basis deformation is chosen as $\beta_{\rm basis}=\beta_{20}$ for $\beta_{20}\le 0.3$ and $\beta_{\rm basis}=\beta_{20}/2$ for $\beta_{20}>0.3$ \cite{Lu2012_PRC85_011301,Lu2014_89_014323}. The densities, scalar potential, and vector potential are calculated on a spatial lattice with 36, 20, and 18 mesh points in the $r_{\perp}$, $\phi$, and $z$ directions, respectively, ensuring the convergence of the calculation results. 

 \begin{figure}[htbp] % [t]表示优先放在页面顶部
    \centering
    \includegraphics[width=0.48\textwidth]{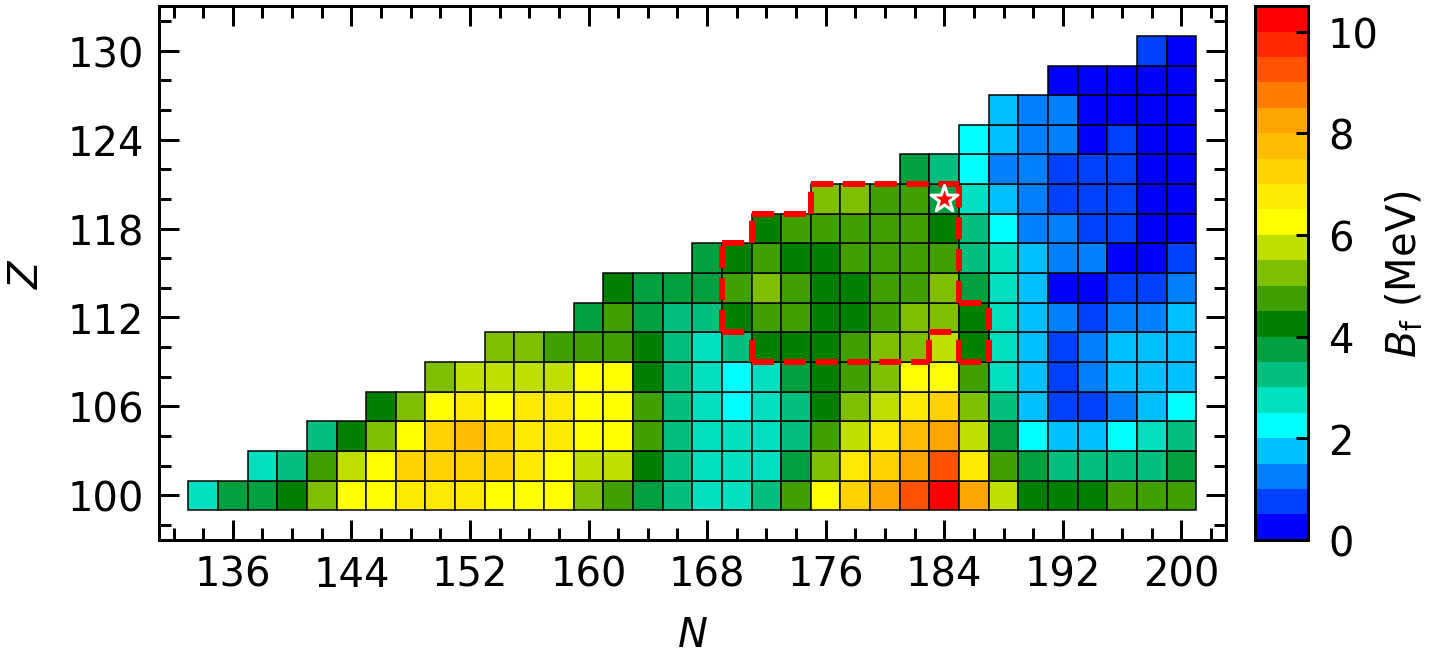} % 使用\textwidth自动适应通栏宽度
    \caption{\justifying Calculated fission barrier heights $B_{\rm{f}}$ of even-even superheavy nuclei. The predicted doubly magic nucleus~${}^{304}120$ is marked with a red star outlined in white. The nuclei around it with $B_{\rm f}=4$--$5.5$ MeV are enclosed by a red dashed contour, highlighting the flatness of the landscape in the  ``island of stability'' of superheavy nuclei.}
    \label{I}
\end{figure}

Figure~\ref{I} shows the calculated fission barrier heights for 292 even-even heavy and superheavy nuclei with $100\le Z\le130$, covering isotopic chains from the two-proton drip line to $N=200$. Local maxima appear at $N = 152$ in Fm, No, and Rf isotopes and at $N=162$ in Hs, Cn, and Ds isotopes, consistent with the known deformed neutron shell gaps \cite{Ghiorso1954_PR95_293,Dvorak2006_PRL_97_242501,Oganessian2013_PRC_87_034605,Hessberger2017_EPJA53_75}. This consistency provides a benchmark for the calculated barrier systematics.

 \begin{figure*}[t] % [t]表示优先放在页面顶部
    \centering
    \includegraphics[width=0.8\textwidth]{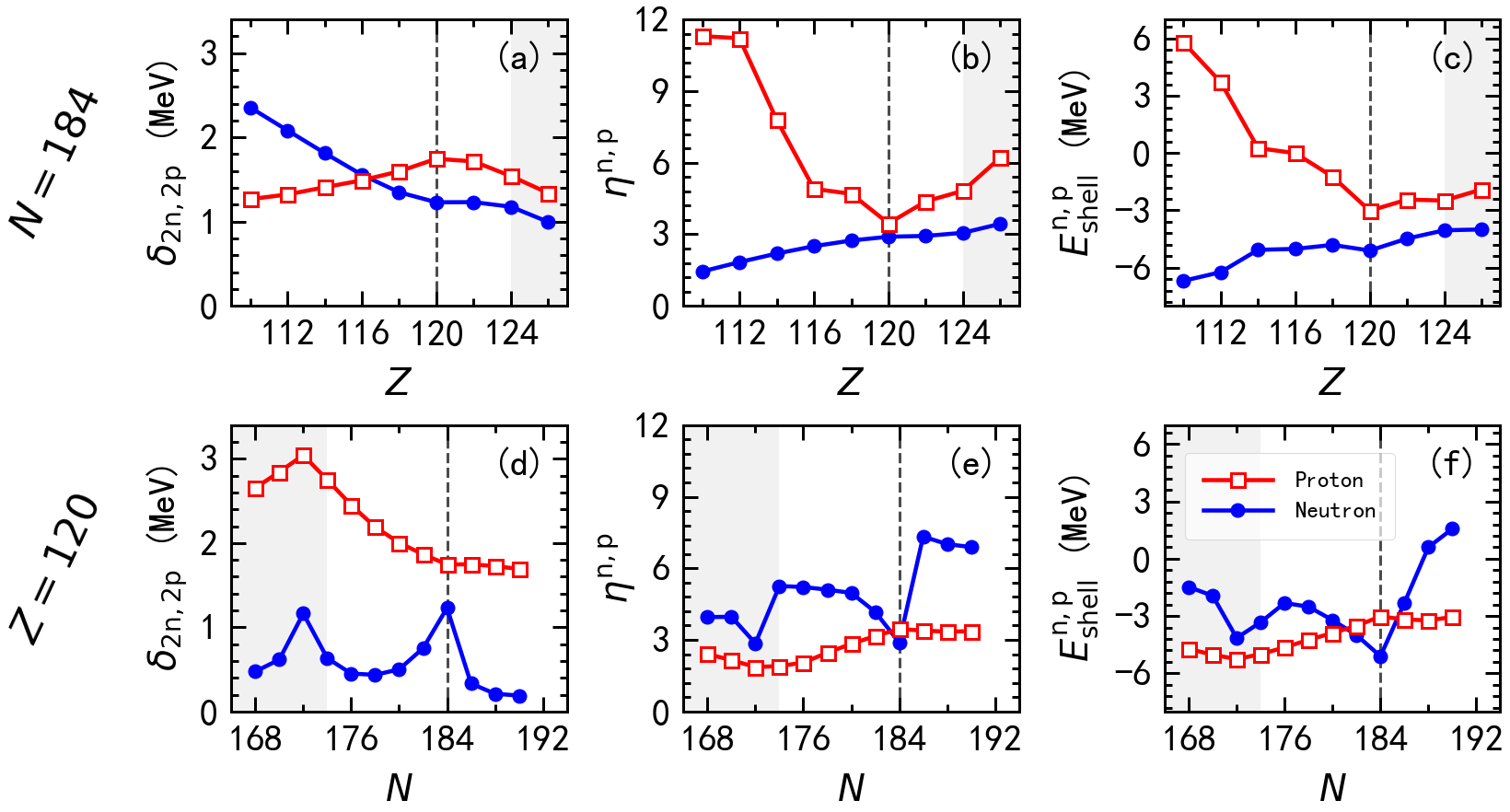} % 使用\textwidth自动适应通栏宽度
         \caption{\justifying Various shell-effect indicators in the $N=184$ isotonic chain~(a)--(c) and the $Z=120$ isotopic chain~(d)--(f). The red line with open squares and the blue line with filled circles represent the corresponding proton and neutron indicators, respectively. The gray dashed lines mark the predicted magic numbers $Z=120$ and $N=184$. The points in the gray shaded region denote nuclei beyond the two-proton drip line.}
    \label{II}
\end{figure*}

    Around the predicted doubly magic nucleus ${}^{304}120$ \cite{ZHANG2005_NPA753_106,ZHANG2022_ADNDT144_101488,GUO2024_ADNDT158_101661,Guo2026}, the calculated fission barrier heights remain at approximately 5 MeV, pointing to a rather flat landscape in the ``island of stability''. In the conventional picture, the shell effects are expected to be strongest near doubly magic nuclei due to MEM \cite{Lunney2003_RMP75_1021,Vladimir2023_EPJA59_22}. In the superheavy region, a strong shell effect generally implies a higher fission barrier, since fission barriers in superheavy nuclei are mainly related to them \cite{Swiatecki1955_PR100_937,Myers1967_AF36_343,STRUTINSKY1967_NPA95_420,NILSSON1969_NPA131_1,MARDYBAN2026_PRC113_054308}. The fission barrier heights are therefore expected to peak around the doubly magic nucleus and gradually decrease as the proton and neutron numbers deviate from the magic numbers, producing a peaked barrier landscape \cite{MAMDOUH2001_NPA679_337,Staszczak2013_PRC87_024320,MOLLER2015_PRC91_024310,JACHIMOWICZ2021_ADNDT138_101393,Malov2021_PRC104_L011304,Wang2024_CPC48_094103}. However, the flat landscape of fission barriers revealed in this work suggests a breakdown of MEM near the predicted doubly magic nucleus. This behavior clearly differs from that observed in known nuclei.

 % To explain this anomaly, we propose the concept of mutually suppressed magicity (MSM), defined as the mutual weakening of the proton and neutron shell closures near the predicted doubly magic nucleus. The MSM would prevent the expected enhanced shell stabilization from a peak in the ``island of stability''. To examine the emergence of the MSM, we employ three indicators to study the shell effects in the ground states as fission barriers in superheavy nuclei are mainly related to them \cite{Swiatecki1955_PR100_937,Myers1967_AF36_343,STRUTINSKY1967_NPA95_420,NILSSON1969_NPA131_1}. The first indicator is the two-proton (neutron) gap $\delta_{\rm{2p (2n)}}$ \cite{Lunney2003_RMP75_1021}:
 To understand the flatness, we employ several indicators to study the shell effects in the ground states. The first indicator is the two-proton (neutron) gap $\delta_{\rm{2p (2n)}}$ \cite{Lunney2003_RMP75_1021}:
\begin{equation}
\begin{aligned}
    \delta_{\rm{2p}}(N,Z)&=S_{\rm{2p}}(N,Z)-S_{\rm{2p}}(N,Z+2)\\
                    &=2B(N,Z)-B(N,Z+2)-B(N,Z-2),\\
    \delta_{\rm{2n}}(N,Z)&=S_{\rm{2n}}(N,Z)-S_{\rm{2n}}(N+2,Z)\\
                    &=2B(N,Z)-B(N+2,Z)-B(N-2,Z),\\
\end{aligned}
\end{equation}
where $S_{\rm 2p}(N,Z)$ and $S_{\rm 2n}(N,Z)$ are the two-proton and two-neutron separation energies, respectively, and $B(N,Z)$ is the binding energy of a nucleus with proton and neutron numbers $Z$ and $N$. The second indicator is the ``energy level density'' $\eta^{\rm{n,p}}$, which is proved highly sensitive to shell effects along the fission path \cite{Bernard2023_EPJA59_51}:  
\begin{equation}
\begin{aligned}
 \eta^{\rm{n,p}}=\sum_k f(\varepsilon_k-\varepsilon^{\rm{n,p}}_0),\\
 f(x)=\begin{cases}
1-\frac{\left|x\right|}{\Delta E}, & \left|x\right|\le\Delta E \\
0,   & \left|x\right|>\Delta E,
\end{cases}
 \end{aligned}
\end{equation}
where $\varepsilon_k$ is the single-particle energy and $\varepsilon^{\rm{n,p}}_0$ are defined as halfway between the two levels surrounding the neutron and proton Fermi energies, respectively. The energy window $\Delta E$ should be of the order of typical shell energy gaps \cite{Bernard2023_EPJA59_51}. Here we choose $\Delta E = 2$ MeV for protons and 1.5 MeV for neutrons. The third indicator is the shell correction energy $E_{\rm{shell}}$ \cite{STRUTINSKY1967_NPA95_420}:
\begin{equation}
    E_{\rm{shell}}=2\sum_{k} n_k \varepsilon_k-2\int^{\tilde{\lambda}}_{-\infty}\tilde{g}(\varepsilon)\varepsilon\rm{d}\varepsilon,
\end{equation}
where $n_k$ represents the occupation probability of each energy level and the smoothed function is defined by
\begin{equation}
    \tilde{g}(\varepsilon)=\frac{1}{\gamma_0\sqrt{\pi}}\sum_ke^{-x^2}\sum_{n=0,2,...}^{6}c_nH_n(x), x=\frac{\varepsilon-\varepsilon_k}{\gamma_0},
\end{equation}
where $c_n=\frac{(-1)^{n/2}}{2^{n}(n/2)!}$ and $H_n$ is the Hermite polynomial of order $n$. The smoothed Fermi energy $\tilde{\lambda}$ is determined by particle-number conservation and the smearing parameter $\gamma_0=1.45 \hbar\omega_0$ is used, where $\hbar\omega_0=41/A^{1/3}$ MeV. All three indicators provide complementary measures of shell effects. Strong shell effects are related to a large two-proton or two-neutron gap $\delta_{\rm 2p(2n)}$, a small ``energy level density'' $\eta^{\rm{n,p}}$, and a large negative shell correction energy $ E_{\rm{shell}}$. 

Figure \ref{II} shows the three shell-effect indicators in the $N=184$ isotonic chain and the $Z=120$ isotopic chain. In the $N=184$ isotonic chain, the two-proton gap $\delta_{\rm{2p}}$ reaches a maximum at $Z=120$. The proton ``energy level density'' $\eta^{\rm{p}}$ reaches a minimum at the same proton number. The proton shell correction energy $E^{\rm{p}}_{\rm{shell}}$ also has a large negative minimum at $Z=120$. These features support $Z=120$ as the next proton magic number. Similarly, the indicators in the $Z=120$ isotopic chain support $N = 184$ as the next neutron magic number. Strong shell effects also occur at $N=172$. However, $^{292}120$ lies beyond the two-proton drip line and its spherical solution is not isotonically robust~\cite{Agbemava2015_PRC_92_054310,ZHANG2022_ADNDT144_101488,GUO2024_ADNDT158_101661}. We therefore do not firmly identify $N=172$ as a neutron magic number. For the candidate proton magic number $Z=114$ \cite{SOBICZEWSKI1966_PL_22_500,Meldner1967_AF36_593,NILSSON1969_NPA131_1,Mo2014_PRC90_024320,JACHIMOWICZ2021_ADNDT138_101393}, the $N=184$ isotonic chain shows no evidence for a strong proton shell effect there (Fig.~\ref{II}). For $^{292}120$ and $^{286}$Fl in the $N=172$ isotonic chain, the total shell correction energies are $-9.4$ and $-6.6$ MeV. Their quadrupole deformations are $\beta_{20}=0$ and 0.154, respectively. These results do not support $Z=114$ as a proton magic number in RMF calculations. 

As shown in Fig.~\ref{II}, in the $N=184$ isotonic chain, the neutron shell effects are not enhanced near $Z=120$ and become weaker as the proton number increases. Similarly, in the $Z=120$ isotopic chain, the proton shell effects at $N = 184$ are not visibly enhanced. This behavior differs from MEM in known doubly magic regions. Instead, as the proton or neutron number approaches the next magic number, the proton and neutron shell closures weaken each other around the predicted doubly magic nucleus. 

We refer to this mutual weakening as ``mutually suppressed magicity'' (MSM).  As the proton or neutron number approaches the next magic number, the Fermi level approaches a shell gap, which would normally strengthen shell effects. However, the calculations show that the shell-closure strength actually becomes weaker. This competition prevents the fission barrier from exhibiting a pronounced maximum near the next doubly magic nucleus, resulting in a flat landscape in the ``island of stability''.

 \begin{figure}[htbp] % [t]表示优先放在页面顶部
    \centering
    \includegraphics[width=0.45\textwidth]{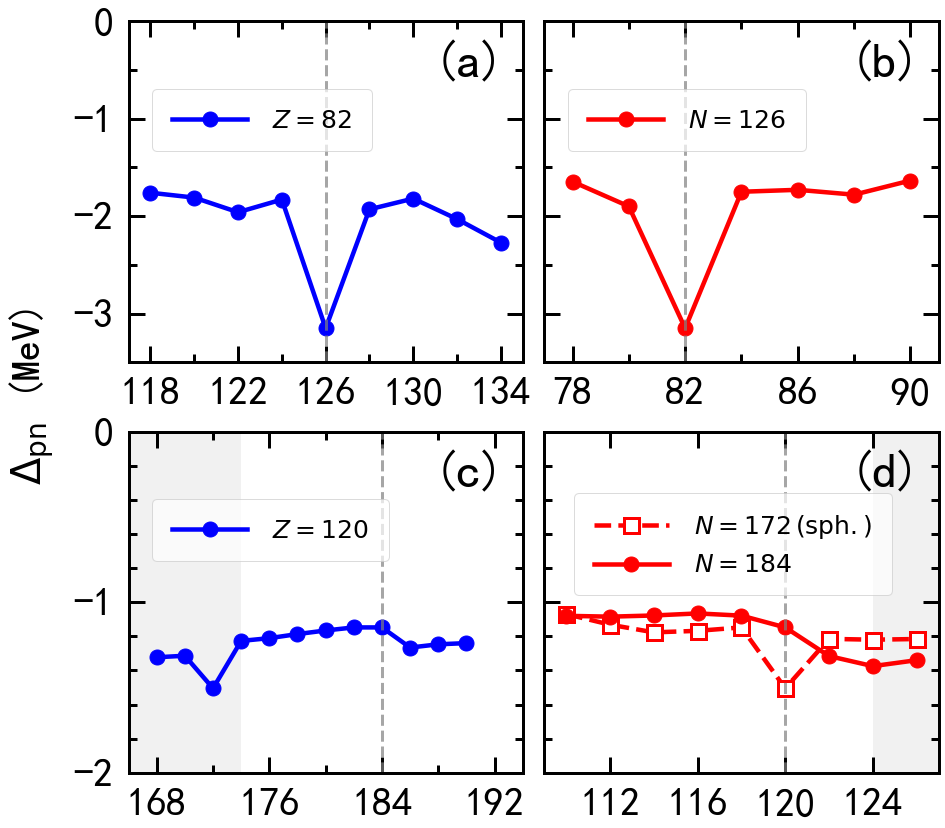} % 使用\textwidth自动适应通栏宽度
    \caption{\justifying $\Delta_{\rm pn}$ for the isotopic chain $Z=82$ (a), the isotonic chain $N=126$ (b), the isotopic chain $Z=120$ (c), and the isotonic chains $N=172$ and $N=184$ (d). The experimental data in panels (a) and (b) are taken from Refs.~\cite{Huang2021_CPC45_030002,Wang2021_CPC45_030003}.}
    \label{VII}
\end{figure}
 \begin{figure*}[t] % [t]表示优先放在页面顶部
    \centering
    \includegraphics[width=0.8\textwidth]{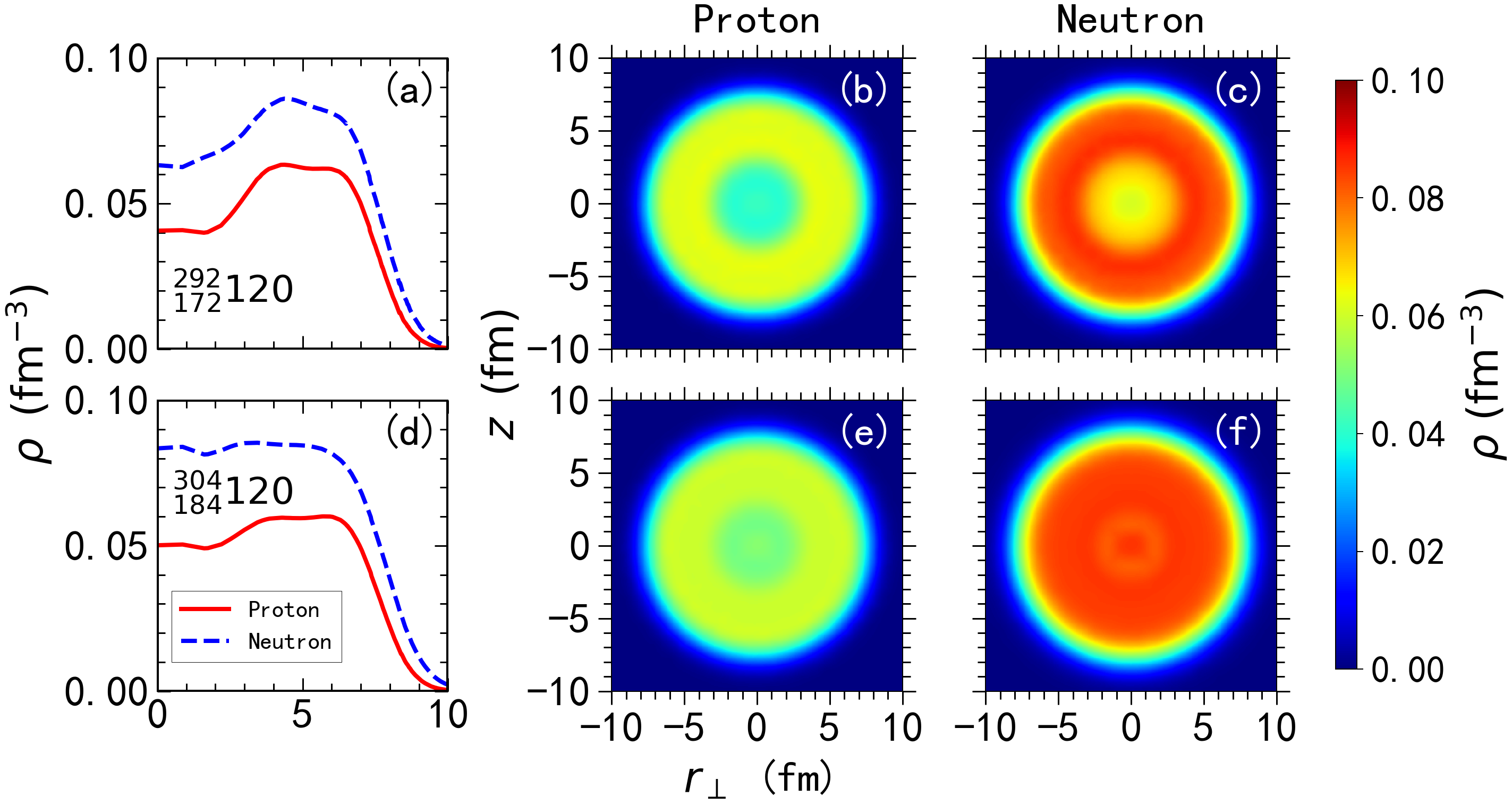} % 使用\textwidth自动适应通栏宽度
    \caption{\justifying Density distributions of ${}^{292}120$ [(a)--(c)] and ${}^{304}120$ [(d)--(f)]. Panels (a) and (d) show the proton and neutron radial density distributions. The red solid and blue dashed lines denote proton and neutron densities, respectively. %Panels (b) and (c) [(e) and (f)] show the density distribution of proton and neutron in the plane $\phi = 0$ of ${}^{292}120$ (${}^{304}120$). 
    Panels (b) and (c) [(e) and (f)] show the proton and neutron density distributions, respectively, in the $\phi=0$ plane for ${}^{292}120$ (${}^{304}120$).
    }
    \label{III}
\end{figure*}

 \begin{figure}[htbp] % [t]表示优先放在页面顶部
    \centering
    \includegraphics[width=0.48\textwidth]{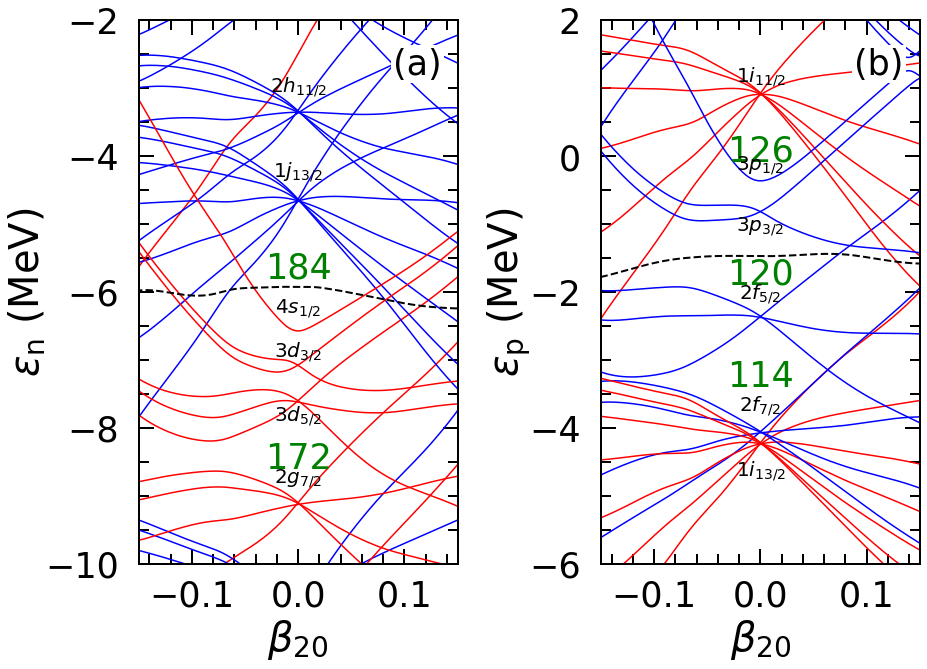} % 使用\textwidth自动适应通栏宽度
    \caption{\justifying Single-neutron~(a) and single-proton~(b) levels of ${}^{304}120$ as functions of quadrupole deformation $\beta_{20}$. The red and blue solid lines represent levels with positive and negative parities, respectively, while the black dashed lines represent the Fermi levels.}
    \label{IV}
\end{figure}

The emergence of MSM reflects an anomalous neutron-proton interaction energy around shell gaps. A widely used indicator of empirical neutron-proton interaction energy is the double binding energy difference $\delta V_{\rm{pn}}$ \cite{ZHANG1989_PLB1_1,BRENNER1990_PLB1_1,Casten2016_ps91_033004}
\begin{equation}
\begin{aligned}
\delta V_{\rm{pn}} = -\frac{1}{4} [&B(N,Z)-B(N-2,Z)\\
                      &-B(N,Z-2)+B(N-2,Z-2)]\\
                   =\,\frac{1}{4} [&S_{\alpha}(N,Z)-S_{2\rm{n}}(N,Z)-S_{2\rm{p}}(N,Z)]+\frac{1}{4}B_\alpha    ,
\end{aligned}
\end{equation}
where $S_{\alpha}(N,Z)=-Q_\alpha(N,Z)$ and $B_\alpha$ is the binding energy of the $\alpha$ particle. Rather sharp jumps in  $\delta V_{\rm{pn}}$ are noted as the proton or neutron number crosses a shell closure \cite{BRENNER1990_PLB1_1,Casten2016_ps91_033004}. These jumps are less pronounced in superheavy nuclei. To characterize the coherent variation of the proton and neutron shell effects across an $\alpha$-linked pair of nuclei, we introduce a more sensitive observable $\alpha$-linked energy difference $\Delta_{\rm pn}$
\begin{equation}
\begin{aligned}
\Delta_{{\rm{pn}}} =&\ 4[\delta V_{\rm{pn}}(N,Z)+\delta V_{\rm{pn}}(N+2,Z+2)]\\
=&\ [\delta_\alpha(N,Z)-\delta_{2\rm{n}}(N,Z)-\delta_{2\rm{p}}(N,Z)],
\end{aligned}
\end{equation}
where $\delta_\alpha(N,Z)=S_{\alpha}(N,Z)-S_{\alpha}(N+2,Z+2)$. By subtracting the separate proton and neutron shell-gap contributions
from the variation of the $\alpha$-separation energy, $\Delta_{\rm{pn}}$
emphasizes the coherent variation of the proton and neutron shell effects. It serves as a more sensitive indicator of MEM or MSM. Around known doubly magic nuclei, such as $^{208}$Pb [Fig.~\ref{VII}~(a) and~(b)], pronounced dips appear at magic numbers. These dips indicate a more attractive proton-neutron shell coupling and provide a signature of MEM. In the superheavy region, as shown in Fig.~\ref{VII} (c) and (d), no comparable dips appear at $N = 184$ along the $Z = 120$ isotopic chain or at $Z = 120$ along the $N = 184$ isotonic chain. Together with the shell indicators in Fig.~\ref{II}, this behavior supports the interpretation that the $Z=120$ and $N=184$ shell closures do not mutually reinforce each other, leading to MSM. For comparison, the spherical neutron shell closure at $N = 172$ is also shown in Fig.~\ref{VII}~(d), where the shapes are constrained to spherical configurations to isolate the effects of deformation. The behavior of the $Z=120,N=172$ combination is closer to that around conventional magic numbers, i.e., MEM.

To understand the microscopic origin of MSM, we examine  the density distributions of $^{292}120$ and $^{304}120$ in Fig.~\ref{III}. Both the proton and neutron density distributions in $^{292}120$ exhibit pronounced central depressions, consistent with a semi-bubble structure. In $^{304}120$, by contrast, only the proton density distribution shows a slight central  depression. The repulsive isovector interaction favors similar neutron and proton density profiles, allowing the spherical shell gaps at $N=172$ and $Z=120$ to reinforce each other in $^{292}120$. For the shell gaps at $N=184$ and $Z=120$, the dissimilar proton and neutron density distributions instead lead to mutually suppressed shell effects. This suppression reduces the shell stabilization expected from the doubly magic nuclei and thereby accounts for the flat landscape in the “island of stability”.

% Furthermore, the central depression density distributions also increase the spin-orbit splitting (e.g. $\pi3p_{1/2}\leftrightarrow\pi3p_{3/2}$, $\pi2f_{5/2}\longleftrightarrow\pi2f_{7/2}$, and $\nu3d_{3/2}\leftrightarrow\nu3d_{5/2}$) and decrease pseudo-spin splitting (e.g. $\pi2f_{5/2}\leftrightarrow\pi3p_{3/2}\leftrightarrow\nu3d_{5/2}$ and $\nu2f_{7/2}$) via affecting the self-consistent potential $V(\vec{\mathbf{r}})$ and $S(\vec{\mathbf{r}})$.                              
Figure~\ref{IV} shows the single-particle levels of $^{304}120$. The formation of bubble structures is usually associated with the depletion of the occupation of $s$ states, which give the dominant contribution to the density at the center of a nucleus \cite{Afanasjev2005_PRC71_024308,Perera2022_PRC106_024321}. The single-neutron levels in Fig.~\ref{IV}~(a) show that the $\nu4s_{1/2}$ orbital is occupied at $N = 184$ but empty at $N = 172$. As a result, the neutron density profiles of $^{304}120$ and $^{292}120$ are characterized by a flat central region and a pronounced central depression, respectively. For the single-proton levels in Fig.~\ref{IV} (b), no unoccupied $s$-wave orbitals are found near the Fermi surface. Instead, vacancies appear in low-angular-momentum orbitals such as $\pi3p_{1/2}$ and $\pi3p_{3/2}$, while high-angular-momentum orbitals such as $\pi2f_{5/2}$ and $\pi2f_{7/2}$ are occupied. The occupations of proton and neutron levels with different angular momenta near the Fermi surfaces in $^{304}120$ give rise to distinct density distributions, which in turn lead to mutually suppressed shell effects at the corresponding shell closures and consequently to a flat landscape in the ``island of stability''.

In summary, we performed the first systematic microscopic calculations of fission barrier heights for even-even superheavy nuclei with triaxial and octupole deformations included. Contrary to the conventional expectation of a peak based on mutually enhanced magicity, the calculated fission barrier heights remain at approximately 5 MeV in the vicinity of the doubly magic nucleus ${}^{304}120$. This indicates an unexpectedly flat landscape in the predicted ``island of stability''. We demonstrate that this flatness is associated with mutually suppressed magicity and propose a new indicator, the $\alpha$-linked energy difference $\Delta_{\rm{pn}}$, to characterize the coherent variation of proton and neutron shell effects. Its behavior around $Z=120$ and $N=184$, together with the conventional shell indicators, shows that the two shell closures do not reinforce each other as they do in known doubly magic nuclei. The distinct  occupations of levels with different angular momenta near the Fermi surface produce dissimilar proton and neutron density distributions, which weaken the corresponding shell closures and prevent the barrier heights from reaching a maximum near ${}^{304}120$, providing a microscopic
interpretation of mutually suppressed magicity. This flat landscape revises understanding of shell evolution in superheavy nuclei and should be taken into account when estimating survival probabilities in heavy-ion fusion-evaporation reactions. It will also be interesting to use the $\alpha$-linked energy difference $\Delta_{\rm{pn}}$ to study bubble nuclei or the properties of new magic numbers, both in superheavy nuclei and in other nuclear regions such as those far from the $\beta$ stability line.

\begin{acknowledgments}

This work is supported by the National Key R\&D Program of China (Grant Nos. 2024YFE0109800 and 2023YFA1606500), the Strategic Priority Research Program of the Chinese Academy of Sciences (Grant Nos. XDB1550100 and XDB0920000), the National Natural Science Foundation of China (Grant Nos. 12447101, 12347139, 12375118, 12435008, 12605207, and W2412043), and the Russian Science Foundation (Grant No. 25-42-00018) \cite{RSF2025}. 
The results described in this paper are obtained on the High-performance Computing Cluster of ITP-CAS and the ScGrid of the Supercomputing Center, Computer Network Information Center of the Chinese Academy of Sciences. 
\end{acknowledgments}


\bibliography{ref.bib}
%\bibliography{xlu.bib,Th229-RZ.bib,th229-xlu.bib}
\end{document}